\documentclass[letterpaper, 10 pt, conference]{ieeeconf}

\IEEEoverridecommandlockouts

\usepackage{graphics}
\usepackage{epsfig}
\usepackage{mathptmx}
\usepackage{times}
\usepackage{amsmath}
\usepackage{amssymb} 

\usepackage{algorithm}%
\usepackage{mathtools}
\usepackage{algpseudocode}%
\usepackage{siunitx}
\usepackage{import}
\usepackage[hidelinks]{hyperref}
\usepackage{booktabs}
\usepackage{multirow}
\usepackage{multirow}

\newtheorem{assumption}{Assumption}
\newtheorem{lemma}{Lemma}
\newtheorem{theorem}{Theorem}
\newtheorem{rem}{Remark}
\newtheorem{res}{Result}
\newtheorem{defn}{Definition}

\usepackage{tikz}
\usepackage{textcomp}
\usepackage{lipsum}

\newcommand\copyrighttext{%
	\footnotesize \textcopyright 2021 IEEE. Personal use of this material is permitted. 
	Permission from IEEE must be obtained for all other uses, in any current or future media, 
	including reprinting/republishing this material for advertising or promotional purposes, 
	creating new collective works, for resale or redistribution to servers or lists, 
	or reuse of any copyrighted component of this work in other works.	
}
\newcommand\copyrightnotice{%
	\begin{tikzpicture}[remember picture,overlay]
		\node[anchor=south,yshift=10pt] at (current page.south) {\fbox{\parbox{\dimexpr\textwidth-\fboxsep-\fboxrule\relax}{\copyrighttext}}};
	\end{tikzpicture}%
}

\title{\LARGE \bf
On the relationship between data-enabled predictive control and subspace predictive control
}

\author{Felix Fiedler$^{1}$  and Sergio Lucia$^{1}$%
	\thanks{*Felix Fiedler acknowledges the support of the Helmholtz Einstein International Berlin Research School in Data Science (HEIBRiDS)}%
	\thanks{$^{1}$ F. Fiedler and S. Lucia are with the Laboratory of Process Automation Systems at Technische Universität Dortmund, Emil-Figge-Str. 70, 44227 Dortmund}
}

\begin{document}

\maketitle
\copyrightnotice
\thispagestyle{empty}
\pagestyle{empty}

\begin{abstract}
Data-enabled predictive control (DeePC) is a recently proposed approach that combines system identification, estimation and control in a single optimization problem, 
for which only recorded input/output data of the examined system is required.
The same premise holds for the subspace predictive control (SPC) method in which a
multi-step prediction model is identified from the same data as required for DeePC. This model is then used to formulate a similar optimal control problem.
In this work we investigate the relationship between DeePC and SPC.
Our primary contribution is to show that SPC is equivalent to DeePC in the deterministic case.
We also show the equivalence of both methods in a special case for the non-deterministic formulation.
We investigate the advantages and shortcomings of DeePC as opposed to SPC with and without measurement noise
and illustrate them with a simulation example.
\end{abstract}

\section{Introduction}
Model predictive control (MPC) is a popular strategy to control multivariable systems with constraints \cite{rawlingsModelPredictiveControl2017}. Traditionally, the main shortcoming of MPC is its computational complexity as it requires the online solution of an optimization problem. Due to advances in computing power, algorithms and efficient implementation strategies, solution time is often not an obstacle for its deployment anymore.

However, MPC still requires a dynamic model of reasonable accuracy to obtain a satisfactory performance. Such model can be obtained from physical modelling, using data-based system identification or a combination of both approaches. Learning from data has recently regained attention although it has been studied for a long time in the field of system identification
\cite{ljungSystemIdentificationTheory1987},
\cite{delchampsStateSpaceInputOutput1988},
\cite{juditskyNonlinearBlackboxModels1995}, \cite{ljungSystemIdentification2017}.
Important categories in this vast field are, among others, linear vs. nonlinear \cite{ljungSystemIdentificationTheory1987} and input/output vs. state-space models \cite{delchampsStateSpaceInputOutput1988}.                
A concrete example of interest  is the autoregressive model with exogeneous inputs (ARX) \cite{ljungSystemIdentification2017}, a popular linear input/output approach.
Typically, ARX models predict the next output of the system and can be evaluated repeatedly to compute finite future trajectories as required for an MPC formulation. Subspace predictive control (SPC) \cite{favoreelSPCSubspacePredictive1999}
follows a similar approach, where a multi-step ahead prediction model is used to compute the finite future trajectory in a single step. 
An extensive overview on the method is given in \cite{huangDynamicModelingPredictive2008} and SPC is still an active field of research with the work in \cite{sedghizadehDatadrivenSubspacePredictive2018}  investigating stability and horizon tuning.

In recent years, a new trend in data-based control seeks to integrate the traditional sequential system identification and control in a unified approach. From the machine learning community, reinforcement-learning is a popular approach following this paradigm \cite{suttonReinforcementLearningSecond2018}. However, reinforcement-learning comes with its own challenges such as the demand for large quantities of data and produces highly variable outcomes \cite{rechtTourReinforcementLearning2019}. 
Another approach to unify identification and control is the newly proposed data-enabled predictive control (DeePC) \cite{coulson2019data} algorithm.
The approach describes a simple configuration of a predictive control problem which operates directly on matrices of collected input/output data. 
Furthermore, the required data has to satisfy only moderate requirements and can often be taken directly from a running process.

The initial work on DeePC \cite{coulson2019data} has thus sparked considerable interest in the control community with several extensions and modifications. The authors in \cite{berberichDataDrivenModelPredictive2020} investigate DeePC with respect to stability and robustness and propose an adaptation to guarantee these properties. Distributionally robust DeePC is presented in \cite{coulson2020distributionally}, where it is assumed that data is sampled from a distribution of possible systems. 
In \cite{alpagoExtendedKalmanFilter2020}, DeePC is combined with an extended Kalman Filter to improve the performance in the case of noisy measurements. 

However, DeePC remains in its nature a linear MPC scheme and is based on data which could also be used for sequential identification and control approaches such as SPC. 
The authors in \cite{coulson2019data} state that an equivalent classical MPC formulation exists for their proposed DeePC scheme.
This equivalence is based on theorizing about the existence of an unknown system parameterization. 
But if such a parameterization could be known, what would be the advantages and potential shortcomings of the DeePC approach, especially in the linear setting, for which DeePC is derived?

The main contribution of this work is to help answer this question. In particular we investigate the relationship between DeePC and SPC.
The performance of SPC and DeePC has recently been empirically compared in \cite{carletDatadrivenPredictiveCurrent2020}. From the obtained similar performance the authors conclude that there are some similarities between the two formulations. 
In this work, we prove that in the linear deterministic case, DeePC is equivalent to SPC. The SPC multi-step prediction model can be determined from the exact same data that DeePC requires to operate. This exact equivalence has, to the best of our knowledge, not yet been established in the literature. 
For the non-deterministic case, DeePC requires some adaptations to deal with the noise corrupted data. We present a minor modification of DeePC and as a second contribution prove that in a special case this formulation is still equivalent to SPC. 
Finally, we compare both controllers numerically on a linear model with additive Gaussian noise. 
We include an analysis of the effect that the amount of available data and the regularization term of DeePC have on the closed-loop performance.

The remainder of this paper is structured as follows. In Section~\ref{sec:DeePC} we present the problem setup and review the data-enabled predictive control algorithm. 
We then present in Section~\ref{sec:notDeePC} the subspace predictive control (SPC) algorithm and prove equivalence to the DeePC method in Theorem~\ref{theo:equivalence}. We proceed to analyse the case of a non-deterministic linear system in Section~\ref{sec:Non_determ_case}. We present a minor modification of the original DeePC problem and prove equivalence to SPC in a special case in Theorem~\ref{theo:non_determ_equivalence}.
Finally, we present a numerical comparison of both controllers in the general case in Section~\ref{sec:Simulation_study}. We finish this work with conluding remarks in Section~\ref{sec:conclusion}.

\section{Data-enabled predictive control}
\label{sec:DeePC}
We investigate an unknown discrete-time LTI dynamic system represented in state-space form:
\begin{subequations}
	\begin{align}
	x_{k+1} &= Ax_k + Bu_k,\\
	y_k &= Cx_k + Du_k,
	\end{align}
\label{eq:generic_LTI_statespace}
\end{subequations}
with $x \in \mathbb{R}^n$ (states), $u\in \mathbb{R}^m$ (inputs), $y\in \mathbb{R}^p$ (measurements) and system matrices $A \in \mathbb{R}^{n\times n}$, $B \in \mathbb{R}^{n\times m}$, $C \in \mathbb{R}^{p\times n}$ and $D \in \mathbb{R}^{p\times m}$.  We assume that the system with $n$ states is given in its minimal representation. 

From this unknown system, we collect $T$ sequences of input/output data of length $L=T_\text{ini}+N$,
where $T_{\text{ini}}$ denotes the length of the initialization sequence and $N$ the length of the prediction horizon.
The collected data is arranged in matrices, such that:
\begin{equation}
\small
	U_L = \left[u^1_L, u^2_L, \dots, u^{T}_L\right], \
	Y_L = \left[y^1_L, y^2_L, \dots, y^{T}_L\right],
\label{eq:U_LY_L_matrix}
\end{equation}
where $u^1_L=\left[\left(u^1_1\right)^\intercal,\dots,\left(u^1_L\right)^\intercal\right]^\intercal\in \mathbb{R}^{Lm}$ and $y^1_L=\left[\left(y^1_1\right)^\intercal,\dots,\left(y^1_L\right)^\intercal\right]^\intercal\in \mathbb{R}^{Lp}$.
These data matrices may or may not coincide with a Hankel or Page matrix (see \cite{coulson2020distributionally}).
We also assume that the sequences were collected starting from $T$ unknown initial states $x_1$:
\begin{equation}
	X_1 = \begin{bmatrix}
		x_1^1, x_1^2,\dots, x_1^T
	\end{bmatrix},
\label{eq:X_1_matrix}
\end{equation}
where $X_1 \in \mathbb{R}^{n\times T}$. 
We divide the input/output data in \eqref{eq:U_LY_L_matrix} in two parts, such that:
\begin{equation}
U_L = \left[
\begin{array}{c}
U_{T_\text{ini}}\\
U_{N}
\end{array}
\right],\ 
Y_L = \left[
\begin{array}{c}
Y_{T_\text{ini}}\\
Y_{N}
\end{array}
\right].
\label{eq:input_output_data}
\end{equation}
To clarify this notation,  we now have $U_{T_\text{ini}} \in \mathbb{R}^{T_{\text{ini}}m\times T}$, $U_{N} \in \mathbb{R}^{Nm\times T}$,
$Y_{T_\text{ini}} \in \mathbb{R}^{T_{\text{ini}}p\times T}$ and $Y_{N} \in \mathbb{R}^{Np\times T}$.
\begin{defn}[\cite{coulson2019data}]
    With $L,T$ such that $T\geq Lm$, the signals comprising matrix $U_L$ are persistently exciting of order $L$ if matrix $U_L$ has full rank.
    \label{defn:persistentlyexciting}
\end{defn}
\begin{defn}[\cite{coulson2019data}]
    We denote with $l(A,B,C,D)$ the lag of system \eqref{eq:generic_LTI_statespace}.
    It is defined as the smallest integer $l$ for which the observability matrix:
    \begin{equation}
        O_l(A,C):=
        \left(
        C,\ CA,\ \dots, CA^{l-1}
        \right),
    \end{equation}
    has full rank.
    \label{def:SystemLag}
\end{defn}
\begin{assumption}
The number of recorded sequences is $T\geq Lm+n$. Matrix $U_L$ is persistently exciting of order $L$ according to Definition~\ref{defn:persistentlyexciting}.
\label{ass:U_L_DeePC}
\end{assumption}
\begin{assumption}
	The initialization sequence is $T_{\text{ini}}>l(A,B,C,D)$, with $l$ according to Definition~\ref{def:SystemLag}.
	\label{ass:lag_larger_T_ini}
\end{assumption}
\begin{assumption}
	The underlying matrix of initial states \eqref{eq:X_1_matrix} has full rank, i.e. $\text{rank}(X_1) = n$
	and the stacked matrix $[X_1^{\intercal}, U_L^{\intercal}]^{\intercal}$ has full rank.

	\label{ass:X_1_full_rank}
\end{assumption}
Subsequently, we state a variant (see Remark~\ref{rem:funamentallemma}) of 
the fundamental lemma of behavioral systems theory \cite{willems2005note}.
\begin{lemma}
Let \eqref{eq:input_output_data} be input/output data of a deterministic system in the form of \eqref{eq:generic_LTI_statespace} which satisfies Assumption~\ref{ass:U_L_DeePC},  Assumption~\ref{ass:lag_larger_T_ini} and Assumption~\ref{ass:X_1_full_rank}.
Any sequence 
$y_{T_\text{ini}}=\left[y_1^\intercal,\dots,y_{T_{\text{ini}}}^\intercal\right]^\intercal$, 
$y_N=\left[y_{T_{\text{ini}}+1}^\intercal,\dots,y_{T_{\text{ini}}+N}^\intercal\right]^\intercal$,
$u_{T_\text{ini}}=\left[u_1^\intercal,\dots,u_{T_{\text{ini}}}^\intercal\right]^\intercal$ 
and
 $u_{N}=\left[u_{T_{\text{ini}}+1}^\intercal,\dots,u_{T_{\text{ini}}+N}^\intercal\right]^\intercal$
is a trajectory of system \eqref{eq:generic_LTI_statespace} if and only if, there exists a $g\in \mathbb{R}^T$, such that:
\begin{equation}
	\left[\begin{array}{c}
	Y_{T_\text{ini}}\\
	U_{T_\text{ini}}\\
	U_{N}\\
	Y_{N}
	\end{array}\right] g 
	= 
	\left[\begin{array}{c}
	y_{T_\text{ini}}\\
	u_{T_\text{ini}}\\
	u_{N}\\
	y_{N}
	\end{array}\right].
	\label{eq:behavioral_system}
\end{equation}
\label{lem:fundamental_lemma}
\end{lemma}
\begin{rem}
    Note that the original statement of the fundamental lemma in \cite{willems2005note} requires the matrices in \eqref{eq:U_LY_L_matrix} to be Hankel matrices. In \cite{coulson2020distributionally} it is shown that \eqref{eq:behavioral_system} also holds for Page matrices which can be extended to show that arbitrary sequences are sufficient as long as the space spanned by the initial states of these sequences has full rank (see Assumption~\ref{ass:X_1_full_rank}). The full proof of Lemma~\ref{lem:fundamental_lemma} is omitted here for brevity. 
    \label{rem:funamentallemma}
\end{rem}

Based on the fundamental Lemma~\ref{rem:funamentallemma}, the authors in \cite{coulson2019data} proposed the elegant data-enabled predictive control scheme (DeePC), where the optimal control policy is obtained as the solution of:
\begin{equation}
	\begin{aligned}
	\min_{g,u_N, y_N}\quad &f(g,u_N, y_N)\\
	\text{s.t.}\quad & \eqref{eq:behavioral_system}\\
	&u_k \in \mathbb{U}\ \forall k\in\{1,\dots,N\},	\\
	&y_k \in \mathbb{Y}\ \forall k\in\{1,\dots,N\}, \\
	\end{aligned}
	\label{eq:DeePC}
\end{equation}

with arbitrary objective function $f(g, u_N, y_N)$.
Within the scope of this work, we only investigate the classical regulation task:
\begin{equation}
    f(g, u_N, y_N) = \sum_{k=1}^N \left(\|y_k\|^2_{Q}+\|u_k\|^2_{R}\right),
    \label{eq:DeePC_costfun}
\end{equation}
where $\|y_k\|_Q^2= y_k^{\intercal} Q y_K$. We require that $Q$ and $R$ are positive definite.

\section{Subspace predictive control}
\label{sec:notDeePC}

We start this section by stating a well known data-based approach for linear system identification: 
The auto-regressive model with external inputs (ARX) \cite{ljungSystemIdentification2017}, where
\begin{equation}
	y_{{T_\text{ini}}+1} = \bar a_1 y_{1}+ \dots + \bar a_{T_\text{ini}} y_{T_\text{ini}}
	+ \bar b_1 u_{1} + \dots + \bar b_{T_\text{ini}} u_{T_\text{ini}}.
	\label{eq:ARX}
\end{equation}
Similarly as in Lemma~\ref{lem:fundamental_lemma}, the equation can only hold if $T_{\text{ini}}>l(A,B,C,D)$, according to Definition~\ref{def:SystemLag} as shown in \cite{markovskyExactApproximateModeling2006}.

To identify the parameters
\begin{equation}
	\bar P = \left[
	\bar a_1 , \ \dots ,\  \bar a_{T_\text{ini}},\  \bar b_1 ,\  \dots ,\ \bar b_{T_\text{ini}}
	\right]
\end{equation}
based on the collected data in \eqref{eq:U_LY_L_matrix},
one would typically solve the least-squares problem:
\begin{equation}
	\min_{\bar P} \quad \|\bar P
	\left[\begin{array}{c}
	Y_{T_\text{ini}}\\
	U_{T_\text{ini}}
	\end{array}
	\right]
	- 
	Y_+
	\|_F^2,
\end{equation}
where $\|\cdot\|_F$ denotes the Frobenius-norm.
The matrix $Y_+$ represents the first $p$ rows of $Y_N$ corresponding to the temporal elements of the output sequences at $T_\text{ini}+1$.
Reminiscent of the ARX model in \eqref{eq:ARX}, the SPC approach \cite{favoreelSPCSubspacePredictive1999} uses the collected data \eqref{eq:U_LY_L_matrix} to directly identify a linear multi-step ahead predictor:
\begin{equation}
\begin{aligned}
	y_N &=  a_1 y_{1} +  \dots +  a_{T_\text{ini}} y_{T_\text{ini}}\\
	& b_1 u_{1}+  \dots +   b_{T_\text{ini}+N} u_{T_\text{ini}+N},
\end{aligned}
\label{eq:multistepahead_prediction}
\end{equation}
where the parameters are again summarized as
\begin{equation}
	P = 
	\left[
	 a_1, \ \dots ,\   a_{T_\text{ini}},\   b_1  ,\  \dots ,\  b_{T_\text{ini}+N} 
	\right].
	\label{eq:P_from_parameters}
\end{equation}
\begin{lemma}
	\label{lem:P_sys_response}
	Let \eqref{eq:input_output_data} be input/output data of a deterministic system in the form of \eqref{eq:generic_LTI_statespace} which satisfies Assumption~\ref{ass:U_L_DeePC},  Assumption~\ref{ass:lag_larger_T_ini} and Assumption~\ref{ass:X_1_full_rank}.
	Let matrix $P^*\in\mathbb{R}^{Np \times Nm+T_{\text{ini}}(m+p)}$ be the solution of  the least squares problem
	\begin{equation}
		P^* = \arg\ \min_{ P} \ \| P
		\underbrace{
			\left[\begin{array}{c}
				Y_{T_\text{ini}}\\
				U_{T_\text{ini}}\\
				U_N
			\end{array}
			\right]}_M
		- 
		Y_N
		\|_F^2,
		\label{eq:p_tilde_ls_problem}
	\end{equation}
	which is expressed explicitly in terms of the Moore-Penrose inverse (denoted with the superscript $\dagger$) as:
	\begin{equation}
		P^* = Y_N M^\dagger.
		\label{eq:p_tilde}	
	\end{equation} 
	Any sequence
	$u_{T_{\text{ini}}}$, $u_N$,
	$y_{T_{\text{ini}}}$,  $y_N$
	is a trajectory of system \eqref{eq:generic_LTI_statespace} if
	\begin{equation}
		y_N =  P^*
		\begin{bmatrix}
		y_{T_\text{ini}}\\
		u_{T_\text{ini}}\\
		u_{N}
		\end{bmatrix}.
		\label{eq:multi_step_ahed_prediction}
	\end{equation}
\end{lemma}
\begin{proof}
	The proof is shown in the appendix.
\end{proof}

Lemma~\ref{lem:P_sys_response} allows to state the subspace predictive control problem (SPC) as
\begin{equation}
	\begin{aligned}
		\min_{u_N,y_N}\quad &\sum_{k=1}^N \left(\|y_k\|^2_{Q}+\|u_k\|^2_{R}\right)\\
		\text{s.t.}\quad & \eqref{eq:multi_step_ahed_prediction}\\
		&u_k \in \mathbb{U},\ \forall k\in \{1,\dots,N\},	\\
		&y_k \in \mathbb{Y},\ \forall k\in \{1,\dots,N\}.	\\
	\end{aligned}
	\label{eq:explicitDeePC}
\end{equation}
\begin{rem}
	In the original work on SPC \cite{favoreelSPCSubspacePredictive1999} the authors choose $T_\text{ini}=N$. For the comparison with DeePC we follow the SPC formulation from \cite{sedghizadehDatadrivenSubspacePredictive2018} with dedicated parameter $T_\text{ini}$.
\end{rem}
\subsection{Equivalence of DeePC and SPC}
In the following we present our main contribution, a theorem stating that subspace predictive control \eqref{eq:explicitDeePC} is an equivalent representation of the DeePC \eqref{eq:DeePC} method.
\begin{theorem}
	Let $U_L$, $Y_L$ in the form of  \eqref{eq:input_output_data}, and 
	$u_{T_{\text{ini}}}$, $y_{T_{\text{ini}}}$ be recorded from 
	the linear system \eqref{eq:generic_LTI_statespace}. The recorded data satisfies 
	 Assumption~\ref{ass:U_L_DeePC},  Assumption~\ref{ass:lag_larger_T_ini} and Assumption~\ref{ass:X_1_full_rank}.
	The optimal control policy $u_N^*$ and optimal output trajectory $y_N^*$ as
	obtained from the solution of the DeePC problem \eqref{eq:DeePC} 
	is equal to the optimal control policy $u_N^*$  and optimal output trajectory $y_N^*$ as obtained from the solution of the SPC  problem  \eqref{eq:explicitDeePC}.
	\label{theo:equivalence}
\end{theorem}
\begin{proof}
We modify problem \eqref{eq:DeePC} and eliminate the variable $g$ from the formulation.
First, we split \eqref{eq:behavioral_system} into: 
\begin{equation}
    \left[\begin{array}{c}
   	Y_{T_\text{ini}}\\
	U_{T_\text{ini}}\\
	U_N
	\end{array}\right] g 
	= 
	\left[\begin{array}{c}
	y_{T_\text{ini}}\\
	u_{T_\text{ini}}\\
	u_N
	\end{array}\right],
	\label{eq:theo_1_01}
\end{equation}
and
\begin{equation}
    y_N = Y_N g.
    \label{eq:y_n=Y_Ng}
\end{equation}
From \eqref{eq:theo_1_01} we can explicitly compute $g^*$ as:
\begin{equation}
    g^* = M^{\dagger}b+\hat g\quad \forall \hat g  \in \ker (M),
    \label{eq:theo_1_g}
\end{equation}
where we denote with $\ker (M)$ the null-space of matrix $M$.
Inserting \eqref{eq:theo_1_g} in \eqref{eq:y_n=Y_Ng} yields:
\begin{align}
    y_N &= Y_N g^*,\\
    \xLeftrightarrow{\eqref{eq:theo_1_g}}
    y_N &= Y_N(M^{\dagger}b+\hat g).
    \label{eq:Theo1_y_N}
\end{align}
We then need to show that $\ker (M) \subseteq \ker (Y_N)$, such that 
\begin{equation}
    Y_N\hat g = 0 \quad \forall \hat g \in \ker (M).
\end{equation}
The relationship $\ker (M) \subseteq \ker (Y_N)$ holds because $P^*$ as obtained from \eqref{eq:p_tilde}, according to Lemma~\ref{lem:P_sys_response}, satisfies \eqref{eq:multi_step_ahed_prediction} and thus:
\begin{equation}
    Y_N = P^*M.
\end{equation}
For any  $x\in \ker (M)$ we have:
\begin{equation}
    Y_Nx = P^*Mx = 0,
\end{equation}
which means $x\in \ker (Y_N)$ and consequently $\ker (M) \subseteq \ker (Y_N)$.
Finally, we obtain from \eqref{eq:Theo1_y_N}:
\begin{align}
    y_N &= Y_N(M^{\dagger}b)\\
    &= P^*b.
\end{align}
We have thus eliminated $g$ from the formulation of \eqref{eq:DeePC}, where the constraint \eqref{eq:behavioral_system} now reads:
\begin{align}
    y_N= P^*\left[\begin{array}{c}
   	y_{T_\text{ini}}\\
	u_{T_\text{ini}}\\
	u_N\\
	\end{array}\right].
\end{align}
This yields the presented SPC formulation \eqref{eq:explicitDeePC} and therefore proves that \eqref{eq:DeePC} and \eqref{eq:explicitDeePC} have the same solution.
\end{proof}
As a consequence of the equivalence between problem \eqref{eq:DeePC} and \eqref{eq:explicitDeePC}, we argue that the DeePC optimization problem implicitly estimates the regressor \eqref{eq:p_tilde} at each control iteration. 
Since the underlying data-matrices \eqref{eq:U_LY_L_matrix}  are collected prior to the control application and are unchanged afterwards, there appears to be no reason for this repeated estimation in the deterministic case. 
The DeePC optimization problem \eqref{eq:DeePC} requires $T+(m+p)N$ optimization variables with $T\geq (T_{\text{ini}}+N)m+n$ (see Assumption \ref{ass:U_L_DeePC}) and $(m+p)(T_{\text{ini}}+N)$ equality constraints. The SPC formulation \eqref{eq:explicitDeePC} has only $(m+p)N$ optimization variables and $pN$ equality constraints.
The increased computational cost of DeePC can be also seen in the numerical example presented in Section~\ref{sec:Simulation_study}.

\section{Non-deterministic case}
\label{sec:Non_determ_case}

We introduce the following linear system subject to zero-mean Gaussian noise $w_k\sim \mathcal{N}(0,\sigma_w^2 I)$:
\begin{subequations}
\begin{align}
	x_{k+1} &= Ax_k + Bu_k,\\
	y_k &= Cx_k + Du_k+w_k.
\end{align}
\label{eq:LTI_noise}
\end{subequations}
As for the deterministic case, data is collected, now resulting in the matrices:
\begin{equation*}
U_L = \left[
\begin{array}{c}
U_{T_\text{ini}}\\
U_{N}
\end{array}
\right],\ 
\tilde Y_L = \left[
\begin{array}{c}
\tilde Y_{T_\text{ini}}\\
\tilde Y_{N}
\end{array}
\right], \ 
\tilde M =   
\left[
\begin{array}{c}
\tilde Y_{T_\text{ini}}\\
U_{T_\text{ini}}\\
U_{N}\\
\end{array}
\right].
\end{equation*}
\subsection{Non-deterministic DeePC}
In the non-deterministic setting, the DeePC formulation \eqref{eq:DeePC} must be adapted because constraint \eqref{eq:behavioral_system} can generally not be satisfied anymore.
These modifications were already proposed in the original work on DeePC \cite{coulson2019data}. 
Following the authors in \cite{berberichDataDrivenModelPredictive2020}, we use $\ell_2$-norms instead of $\ell_1$-norms to penalize the slack variables in the cost function.
We furthermore propose to add an additional slack variable $\sigma_{u}$ which, similarly to $\sigma_{y}$, is motivated by additive input noise. 
This yields the following adapted DeePC formulation:
\begin{equation}
\small
	\begin{aligned}
	\min_{
		g,u_N, y_N, \sigma_{y}, \sigma_{u}
	}\quad &
	\sum_{k=1}^N \left(\|y_k\|^2_{Q}+\|u_k\|^2_{R}\right)\\
	&+\lambda_g \|g\|_2^2+\lambda_{\sigma_y}\|\sigma_{y}\|_2^2 +\lambda_{\sigma_u}\|\sigma_{u}\|_2^2\\
	\text{s.t.}\quad & 
	\left[\begin{array}{c}
	\tilde Y_{T_\text{ini}}\\
	U_{T_\text{ini}}\\
	U_{N}\\
	\tilde Y_N
	\end{array}\right] g 
	= 
	\left[\begin{array}{c}
	y_{T_\text{ini}}\\
	u_{T_\text{ini}}\\
	u_{N}\\
	y_N
	\end{array}\right]
	+
	\left[\begin{array}{c}
	\sigma_{y}\\
	\sigma_{u}\\
	0\\
	0
	\end{array}\right]
	,\\
	&u_k \in \mathbb{U}\ \forall k\in\{1,\dots,N\},	\\
	&y_k \in \mathbb{Y}\ \forall k\in\{1,\dots,N\}. \\
	\end{aligned}
	\label{eq:DeePC_noise}
\end{equation}
\subsection{Non-deterministic SPC}
Subspace predictive control in the non-deterministic setting requires no modifications for the identification step:
\begin{equation}
    \tilde P^* = \tilde Y_N \tilde M^{\dagger}.
    \label{eq:p_tilde_noise}
\end{equation}
The SPC optimization problem is adapted to account for noisy initial conditions with slack variables $\sigma_{y}$ and $\sigma_{u}$.
\begin{equation}
	\small
	\begin{aligned}
		\min_{u_N,y_N, \sigma_{y}, \sigma_{u}}\quad &\sum_{k=1}^N \left(\|y_k\|^2_{Q}+\|u_k\|^2_{R}\right)
		+\lambda_{\sigma_y}\|\sigma_{y}\|_2^2 +\lambda_{\sigma_u}\|\sigma_{u}\|_2^2\\
		\text{s.t.}\quad &
		y_N = 
		\tilde P^*
		\left(
		\begin{bmatrix}
			y_{T_{\text{ini}}}\\
			u_{T_{\text{ini}}}\\
			u_{N},
		\end{bmatrix}+
		\begin{bmatrix}
			\sigma_y\\
			\sigma_u\\
			0,
		\end{bmatrix}
		\right)
		\\
		&u_k \in \mathbb{U},\ \forall k\in \{1,\dots,N\},	\\
		&y_k \in \mathbb{Y},\ \forall k\in \{1,\dots,N\}.	\\
	\end{aligned}
	\label{eq:explicitDeePC_noise}
\end{equation}
The main difference in the non-deterministic setting is that $\tilde P^*$ now cannot exactly map $\tilde M$ onto $\tilde Y_N$. 
However, due to the underlying least-squares formulation \eqref{eq:p_tilde_ls_problem}, the obtained regressor has favorable statistical properties in this setting, such as being unbiased \cite{Verhaegen2007}.
\subsection{Equivalence in the non-deterministic case}
Equivalence of the adapted DeePC problem \eqref{eq:DeePC_noise} and SPC problem \eqref{eq:explicitDeePC_noise} in the non-deterministic case cannot generally be established. 
There are, however, special cases in which both methods yield the exact same optimal control policy $u_N^*$ and optimal output trajectory $y_N^*$. 
Subsequently, we present a theorem to show equivalence in such a special case. 
For this purpose, we reformulate the adapted DeePC problem \eqref{eq:DeePC_noise} and SPC problem \eqref{eq:explicitDeePC_noise}. 
We stack the optimization variables 
$v = \left[
\sigma_{y},\ \sigma_{u}, \ u_N
\right]$
and introduce the block-diagonal weighting matrices:
\begin{equation*}
	\scriptsize
	\tilde Q = 
	\begin{bmatrix}
		Q& & \\
		& \ddots &\\
		&& Q
	\end{bmatrix},\ 
	\tilde R = 
	\begin{bmatrix}
		R& & \\
		& \ddots &\\
		&& R
	\end{bmatrix},\ 
	V = 
	\left[
	\begin{array}{ccc}
		\lambda_{\sigma_y}& & \\
		& \lambda_{\sigma_u} &\\
		& & \tilde R 
	\end{array}
	\right].
\end{equation*}
We then drop the inequality constraints in  \eqref{eq:DeePC_noise} and \eqref{eq:SPC_noise_alt}.
This allows to state the unconstrained version of the adapted DeePC problem \eqref{eq:DeePC_noise} as:
\begin{equation}
	\begin{aligned}
		\min_{g,v, y_N}\quad & \frac{1}{2}g^{\intercal} (\lambda_g I) g + \frac{1}{2}y_N^{\intercal} \tilde Q y_N + \frac{1}{2}v^{\intercal} V v\\
		\text{s.t.}\quad &
		y_N = \tilde Y_N g\\ 
		&\underbrace{
			\left[\begin{array}{c}
				\tilde Y_{T_\text{ini}}\\
				U_{T_\text{ini}}\\
				U_{N}
			\end{array}\right]}_{\tilde M} g 
		= 
		\underbrace{
		\left[\begin{array}{c}
			y_{T_\text{ini}}\\
			u_{T_\text{ini}}\\
			0
		\end{array}\right]}_b
		+
		\underbrace{
			\left[\begin{array}{c}
				\sigma_{y}\\
				\sigma_{u}\\
				u_{N}
			\end{array}\right]}_{v},
	\end{aligned}
	\label{eq:DeePC_noise_alt}
\end{equation}
and the unconstrained version of the SPC problem \eqref{eq:explicitDeePC_noise} as:
\begin{equation}
	\begin{aligned}
	\min_{v,y_N}\quad & \frac{1}{2}y_N^{\intercal}\tilde Q y_N + \frac{1}{2}v^{\intercal} V ^{\intercal}v\\ 
	\text{s.t.}\quad & 
	y_N
	= 
	\tilde P
	\left(
	\underbrace{
	\left[\begin{array}{c}
	y_{T_\text{ini}}\\
	u_{T_\text{ini}}\\
	0
	\end{array}\right]}_b
	+
	\underbrace{
	\left[\begin{array}{c}
	\sigma_{y}\\
	\sigma_{u}\\
	u_{N}\\
	\end{array}\right]}_{v}
	\right).
	\end{aligned}\\
	\label{eq:SPC_noise_alt}
\end{equation}
Dropping the inequality constraints in \eqref{eq:DeePC_noise_alt} and \eqref{eq:SPC_noise_alt} has been considered previously for DeePC in \cite{carletDatadrivenPredictiveCurrent2020} and SPC in \cite{sedghizadehDatadrivenSubspacePredictive2018} 
as it allows for explicit solutions of  \eqref{eq:DeePC_noise_alt} and \eqref{eq:SPC_noise_alt} which are preferable for very fast control applications.
Below, we state these explicit solutions in our notation.
\begin{res}
	\label{res:v_opt_DeePC}
	The optimal value $v^*$ for \eqref{eq:DeePC_noise_alt} is computed explicitly as:
	\begin{equation}
		\label{eq:v_opt_DeePC}
		v^* = \tilde M(\lambda_g I+\tilde M^{\intercal}V \tilde M+Y_N^{\intercal}\tilde QY_N)^{-1}\tilde M^{\intercal}Vb-b.
	\end{equation}
\end{res}
\begin{proof}
	The proof is shown in the appendix.
\end{proof}
\begin{res}
	\label{res:v_opt_SPC}
	The optimal value $v^*$ for \eqref{eq:SPC_noise_alt} is computed explicitly as:
	\begin{equation}
		\label{eq:v_opt_SPC}
		v^* = -(V+\tilde P^{\intercal}Q \tilde P)^{-1}\tilde P^{\intercal} \tilde Q \tilde Pb.
	\end{equation}
\end{res}
\begin{proof}
	The proof is shown in the appendix.
\end{proof}
We use Result~\ref{res:v_opt_DeePC} and Result~\ref{res:v_opt_SPC} to establish equivalence of the DeePC and SPC solution $v^*$ under certain conditions.
\begin{theorem}
	\label{theo:non_determ_equivalence}
	Let $T=Lm+T_{\text{ini}}p$ and Assumption~\ref{ass:U_L_DeePC}, Assumption~\ref{ass:lag_larger_T_ini} 
	and Assumption~\ref{ass:X_1_full_rank}
	hold. 
	Let $M\in\mathbb{R}^{Lm+T_{\text{ini}}p\times T}$
	and $Y_N\in\mathbb{R}^{Np\times T}$ have full rank.
	The unconstrained DeePC problem \eqref{eq:DeePC_noise_alt} and unconstrained SPC problem \eqref{eq:SPC_noise_alt}
	yield the same optimal solution $v^*$, if $\lambda_g = 0$. 
\end{theorem}
\begin{proof}
	We reformulate \eqref{eq:v_opt_SPC} by applying the following identity \cite{hendersonDerivingInverseSum1981}:
	\begin{equation}
		\label{eq:matrix_inversion_lemma_01}
		(\Lambda+\Gamma)^{-1}=\Lambda^{-1}-(\Lambda+\Gamma)^{-1}\Gamma \Lambda^{-1}
	\end{equation}
	with
	\begin{equation}
		\Lambda = \tilde P^{\intercal}\tilde Q \tilde P,\quad \Gamma = V.
	\end{equation}
	Note that \eqref{eq:matrix_inversion_lemma_01} only holds if $\Lambda$ and $\Lambda + \Gamma$ are non-singular. 
	According to the assumptions, we have that $\tilde M$ and $\tilde Y_N$ have full rank and $\tilde M$ is quadratic,
	such that:
	\begin{equation}
		\tilde P^* = Y_N\tilde M^{\dagger}  
		= \tilde Y_N \tilde M^{-1} 
		\label{eq:theo_2_proof_P}
	\end{equation}
	also has full rank. By definition $Q$, $R$ and therefore $\tilde Q$ and $V$ are positive definite.
	Therefore we have that $\tilde P^{\intercal}\tilde Q \tilde P$  
	and $\tilde P^{\intercal}\tilde Q \tilde P+V$
	are positive definite. 
	Identity \eqref{eq:matrix_inversion_lemma_01} can thus be applied 
	to the explicit solution of the SPC problem \eqref{eq:v_opt_SPC} and yields:
	\begin{equation}
		\begin{aligned}
		v^* &= \left[-(\tilde P^{\intercal}\tilde Q\tilde P)^{-1} +(V+\tilde P^{\intercal}\tilde Q \tilde P)^{-1}V(\tilde P^{\intercal}\tilde Q \tilde P)^{-1}\right]\tilde P^{\intercal}\tilde Q\tilde P b\\
		&= (V+\tilde P^{\intercal}\tilde Q \tilde P)^{-1}Vb -b.
		\end{aligned}
		\label{eq:theo_2_proof_v_opt}
	\end{equation}
	We substitute \eqref{eq:theo_2_proof_P} in \eqref{eq:theo_2_proof_v_opt}:
	\begin{equation}
	\begin{aligned}
		v^*&= \left(V+(\tilde M^{-1} )^{\intercal}\tilde Y_N^{\intercal} \tilde Q \tilde Y_N \tilde M^{-1} \right)^{-1}Vb-b\\
		&= \left(\left(\tilde M^{-1} \right)^{\intercal} (\tilde M^{\intercal} V \tilde M +\tilde Y_N^{\intercal}  \tilde Q \tilde Y_N) \left(\tilde M^{-1}\right) \right)^{-1}  Vb-b,
	\end{aligned}
	\end{equation}
	and computing the outer inverse, we obtain

	\begin{equation}
		v^*= \tilde M \left(\tilde M^{\intercal} V \tilde M +\tilde Y_N^{\intercal}  \tilde Q \tilde Y_N  \right)^{-1} \tilde M^{\intercal} Vb-b.
		\label{eq:theo_2_proof_v_opt_final}
	\end{equation}
	This means that if $\lambda_g=0$ as required in the theorem, 
	the explicit solution \eqref{eq:v_opt_DeePC} of the unconstrained adapted DeePC problem \eqref{eq:DeePC_noise}
	is equal to \eqref{eq:theo_2_proof_v_opt_final} 
	and therefore equal to the explicit solution \eqref{eq:v_opt_SPC} 
	of the unconstrained SPC problem \eqref{eq:explicitDeePC_noise}.
\end{proof}
Clearly the equivalence established in Theorem~\ref{theo:non_determ_equivalence} holds only under strong assumptions
(i.e. $T=Lm+T_{\text{ini}}p$ and $\lambda_g=0$)
and is shown here only for the unconstrained formulation of DeePC \eqref{eq:DeePC_noise_alt} and SPC \eqref{eq:SPC_noise_alt}.
Furthermore, the presented DeePC formulation in \eqref{eq:DeePC_noise} and \eqref{eq:DeePC_noise_alt} 
deviates from  formulations in previous works through the addition of the slack variable $\sigma_{u}$.
In future work, we therefore aim to investigate the constrained case which would also allow to enforce $\sigma_{u}=0$.

Bearing in mind the limitations of Theorem~\ref{theo:non_determ_equivalence}, we still argue that it shows that DeePC and SPC are very closely related, even in the non-deterministic formulation.
This close relationship also holds in practice as shown in \cite{carletDatadrivenPredictiveCurrent2020} and will be further illustrated in Section~\ref{sec:Simulation_study}.
The main reason for deviations between the DeePC and SPC solution in the general case is due to the regularization of the decision variable $g$ in DeePC with $\lambda_g>0$.
It becomes clear, by inspecting \eqref{eq:v_opt_DeePC}, that this regularization is required in the general case, 
as $\tilde M^{\intercal}V\tilde M+\tilde Y_N^{\intercal}\tilde Q\tilde Y_N$ is singular for 
$T>\max(Lm+T_{\text{ini}}p,\ Np)$.
In Section~\ref{sec:Simulation_study} we numerically investigate the effect of $\lambda_g$ on the obtained solution. 

\section{Simulation study}
\label{sec:Simulation_study}
\begin{figure}
	\footnotesize
	\centering
	\def\svgwidth{0.5\linewidth}
	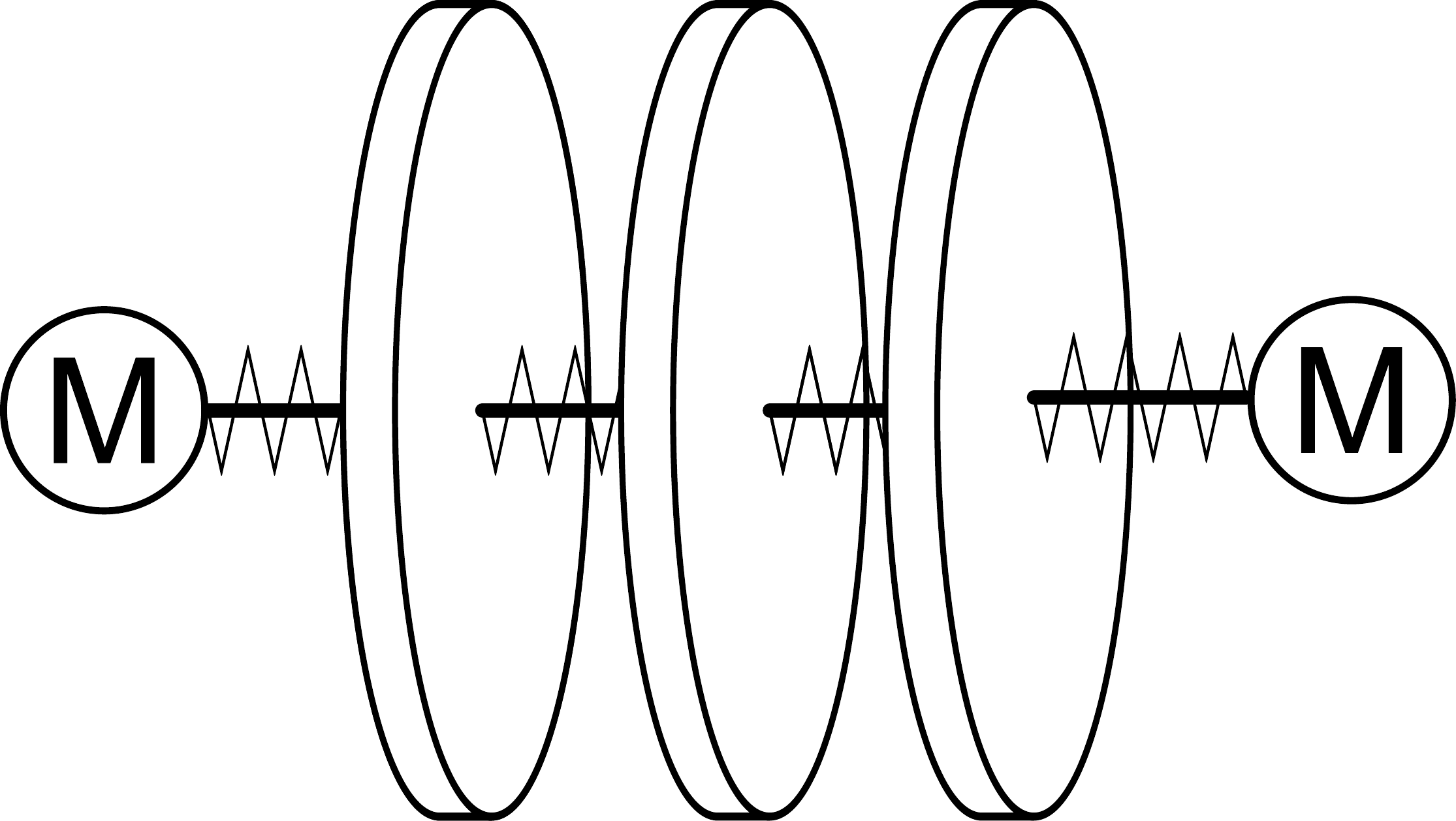
	\caption{Triple-mass-spring system (rotating) with two stepper motors as inputs (motor angle). Disc angles can be measured.}
	\label{fig:triple_mass_spring}
\end{figure}

In this section, we present numerical evidence of our findings. We investigate a simple LTI system in the form of \eqref{eq:generic_LTI_statespace} in two scenarios: with and without additive Gaussian measurement noise. The investigated system, displayed in Figure~\ref{fig:triple_mass_spring}, is a triple-mass-spring system (rotating discs) with two stepper motors ($m=2$) attached to the outermost discs via additional springs. These disc angles are the measured output of the system ($p=3$). The system has a total of $n=8$ states. All materials to reproduce the results in this section (as well as the investigated system) are available in our accompanying repository\footnote{\url{https://github.com/4flixt/DeePC_Perspective}}.
According to Definition~\ref{def:SystemLag} we determine the system lag $l=2$. To satisfy Assumption~\ref{ass:lag_larger_T_ini}, we choose $T_{\text{ini}}=4$.
The prediction horizon is chosen as $N = 40$, which represents \si{4s}.
This means we must collect a minimum of $T>Lm+n=96$ sequences for our data matrices in \eqref{eq:U_LY_L_matrix} to satisfy Assumption~\ref{ass:U_L_DeePC}. 
Note that data is collected from independent experiments with random initial states
fulfilling Assumption~\ref{ass:X_1_full_rank}.
The regulation objective is parameterized with $Q=I$ and $R=0.1 I$ and we constrain the inputs to $\mathbb{U}=\{-0.7,0.7\}$.
All controller variants are implemented using CasADi \cite{Andersson2018} with IPOPT \cite{Andreas2006}.
\subsection{Deterministic system}
We first consider the described system without measurement noise and 
compare the deterministic DeePC algorithm \eqref{eq:DeePC} with SPC \eqref{eq:explicitDeePC}.
The results are obtained with data from $T=150$ captured sequences. 
In the deterministic setting, we have shown in Theorem~\ref{theo:equivalence} that both methods yield identical solutions.
\begin{figure}
	\scriptsize
	\centering
	\def\svgwidth{1\linewidth}
	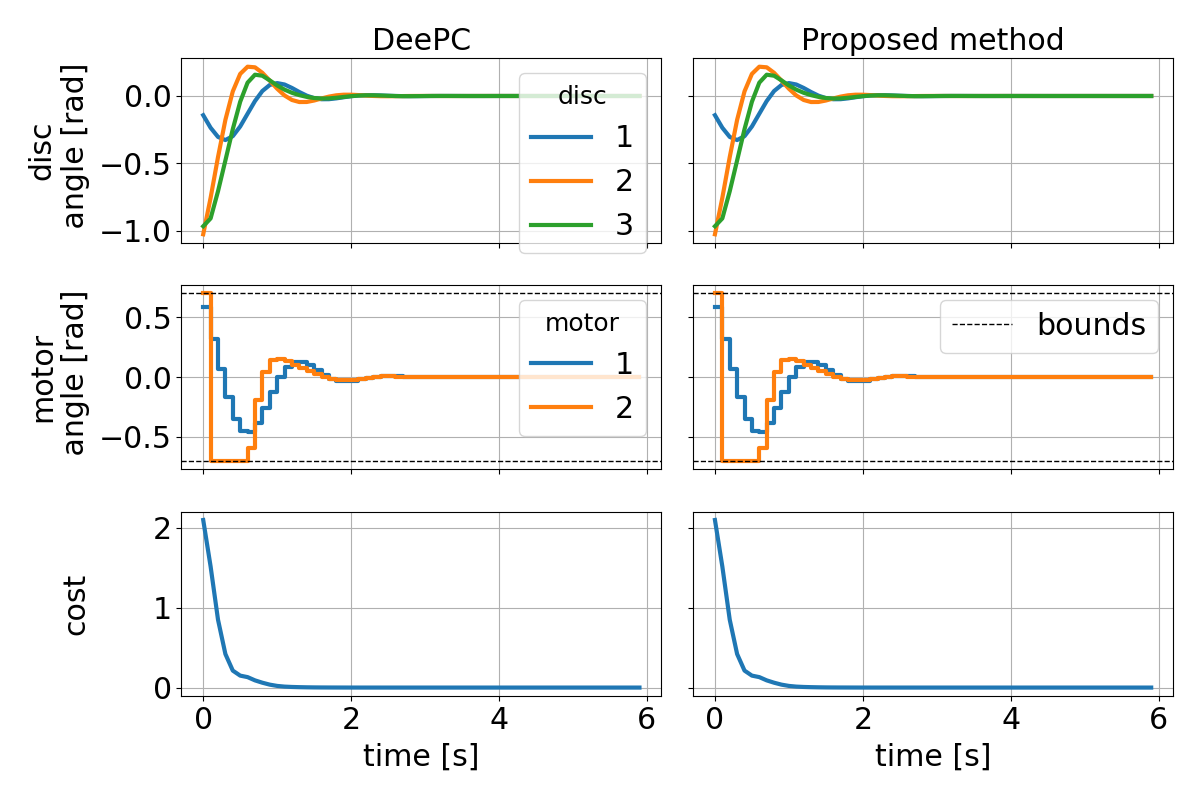
	\caption{Closed-loop trajectories of the mass-spring-system from Figure~\ref{fig:triple_mass_spring}. Regulation after some initial excitation. Comparison of DeePC \eqref{eq:DeePC} and SPC \eqref{eq:explicitDeePC} with deterministic data.}
	\label{fig:01_determ_timeseries_compare}
\end{figure}
In Figure~\ref{fig:01_determ_timeseries_compare} we can see that the identical behavior also appears in practice.
We present  here an exemplary closed-loop trajectory for both methods with random (but identical) excitation phase to obtain $y_{T_{\text{ini}}}$ and $u_{T_{\text{ini}}}$. 
To further quantify the similarity, we compare the cumulated cost:
\begin{equation}
    c = \sum_{k=1}^{N_\text{sim}}\left(
    y_k^{\intercal}Qy_k + u_k^{\intercal}Ru_k
    \right).
    \label{eq:cumulated_cost_simresults}
\end{equation}
For the results in Figure~\ref{fig:01_determ_timeseries_compare} the cumulative cost differs between both controller variants on the order of $10^{-11}$ and equals to $5.617$, 
which will serve as a benchmark for the non-deterministic case. 

\subsection{Non-deterministic system}
In this subsection we investigate the previously described LTI system in the form of \eqref{eq:LTI_noise}, where we choose $\sigma_w = 10^{-2}$ in all experiments. 
We choose $\lambda_{y}=10^{4}$ and $\lambda_{u}=10^4$ for non-deterministic formulations of DeePC \eqref{eq:DeePC_noise} and SPC \eqref{eq:explicitDeePC_noise}. 

We want to compare the performance of both approaches and study the effect 
of the tuning parameter $\lambda_g$ and of the number of recorded sequences $T$ on the cost, 
according to \eqref{eq:cumulated_cost_simresults}, and computation time.

\begin{figure}
	\scriptsize
	\centering
	\def\svgwidth{1\linewidth}
	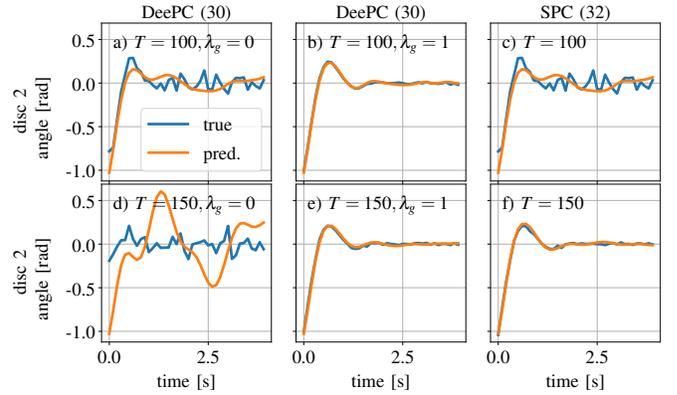
	\caption{Open-loop trajectories for the non-deterministic case with DeePC \eqref{eq:DeePC_noise} vs. SPC \eqref{eq:explicitDeePC_noise}. Comparison of predicted vs. true system response (when subjected to the optimal control trajectory). Only disc 2 (see Figure~\ref{fig:01_determ_timeseries_compare}) is displayed for clarity.}
	\label{fig:02_noise_openloop_pred}
\end{figure}

In Figure~\ref{fig:02_noise_openloop_pred} we present predicted optimal trajectories obtained with \eqref{eq:DeePC_noise} and \eqref{eq:explicitDeePC_noise} for the same initial input/output data $\tilde y_{T_{\text{ini}}}$ and $u_{T_{\text{ini}}}$ (noise disturbed). 
These trajectories (denoted as pred. in Figure~\ref{fig:01_determ_timeseries_compare}) are compared with the true system response resulting from the sequence of optimal inputs.
We vary the number of recorded sequences $T=\{100,150\}$ and change the value of $\lambda_g=\{0,1\}$. For a more concise representation we only showcase the output of the second disc angle. 
Notice that $T=100$ barely exceeds $T=96$, the lower bound to satisfy Assumption~\ref{ass:U_L_DeePC}. 
With $T=100$ and $\lambda_g=0$ we also satisfy the assumptions for Theorem~\ref{theo:non_determ_equivalence}, 
meaning that we expect identical solutions from the unconstrained DeePC problem \eqref{eq:DeePC_noise_alt} and the SPC problem \eqref{eq:SPC_noise_alt}.
This equivalence can also be observed for the constrained DeePC and SPC formulations in Figure~\ref{fig:02_noise_openloop_pred} when comparing case a) and c). 
The open loop-cost of these cases differs on the order of  $10^{-13}$.
Another observation from Figure~\ref{fig:02_noise_openloop_pred} case d) is the unsatisfactory performance of DeePC, 
which stems from the fact that with $\lambda_g=0$ and $T>\max(Lm+T_{\text{ini}}p,\ Np)=120$ a singular matrix arises in the computation of $v^*$.
Good performance with SPC can be obtained with $T=150$ as seen in case f) or with DeePC for $T=150$ and $\lambda_g = 1$ as shown in case e).

\begin{center}
\begin{table}
    \centering
    \caption{Closed-loop trajectories for the non-deterministic case with DeePC \eqref{eq:DeePC_noise} vs. SPC \eqref{eq:explicitDeePC_noise}.}
    \label{tab:closed_loop_cost}
    \begin{tabular}{llcccc}
    & & \multicolumn{3}{c}{\textbf{Number of sequences $T$} }& \textbf{Ref.}\\
    \cmidrule(l){3-5}
    & & 100 & 150 & 200 &\\
    \midrule
    \textbf{DeePC}
    &cost \eqref{eq:cumulated_cost_simresults}& 5.924 & 5.674 & 5.662& 5.617\\
    ($\lambda_g = 1$)&comp. time [ms]& 25.272 & 39.419 & 59.291& -\\
    \midrule
    \multirow{2}{*}{\textbf{SPC}}
    &cost \eqref{eq:cumulated_cost_simresults}& 5.810 & 5.64 & 5.632 & 5.617\\
    &comp. time [ms]& 17.30 & 16.25 & 16.13& -\\
    \bottomrule
    \end{tabular}
\end{table}
\end{center}
A further comparison is presented in Table~\ref{tab:closed_loop_cost}, where we analyze closed-loop trajectories obtained with DeePC \eqref{eq:DeePC_noise}, with $\lambda_g=1$, and SPC \eqref{eq:explicitDeePC_noise}. 
All results in this table are averages over ten simulation experiments, with independently sampled data matrices and measurement noise. 

We take two main conclusions from the presented data in Table~\ref{tab:closed_loop_cost}. 
First, we see that SPC outperforms DeePC both in terms of overall cost and computation time in all scenarios. 
While the advantage of SPC with respect to the cost is minor, the difference in computation time is significant. 
Especially for the case $T=150$, which seems to be a good compromise between performance and cost for DeePC and SPC, we notice a significant increase in computation time. 
Note that we compare only the online CPU time. The computation of the pseudo inverse is excluded, as it is performed offline and only once.
As a second conclusion, we notice that even with $T=100$, DeePC is outperformed by SPC. 
This is counter intuitive in comparison to Figure~\ref{fig:02_noise_openloop_pred} case b) and c), where open-loop trajectories seem better with DeePC. 
However, open-loop predictions are not the same as close-loop control and we find that SPC shows less oscillatory behavior around the origin (not shown here), thus leading to a slightly lower closed-loop cost. 

Note that we also investigated DeePC \eqref{eq:DeePC_noise}, with $\lambda_g\in \{0.1, 10\}$ and $T=150$. 
Compared to the displayed case $\lambda_g = 1$ and $T=150$ in Table~\ref{tab:closed_loop_cost},
the different parameters  for $\lambda_g$ lead to increased closed-loop costs, 
according to \eqref{eq:cumulated_cost_simresults}, of 5.707 and 5.984. 
As a conclusion, we found that $\lambda_g = 1$ is an appropriate choice for the presented comparison. 

\section{Conclusions}
\label{sec:conclusion}
In this work we investigated the relationship between data-enabled predictive control (DeePC)
and subspace predictive control (SPC).
Both methods require for their formulation the exact same data matrices, satisfying identical requirements.
With DeePC, these matrices are directly incorporated in the optimal control problem, 
whereas SPC requires the data to identify a multi-step prediction model.
We show that in the case of deterministic data, both formulations are equivalent.
From the equivalence, we reason that DeePC implicitly estimates the same multi-step ahead prediction model at each iteration, thus adding significant additional computational cost to its online application. 
In particular, the online computational cost of DeePC grows with the number of collected data sequences.

For the non-deterministic case, the exact equivalence between DeePC and SPC does not hold in the general case. 
We propose a minor modification of DeePC and prove that this formulation is equivalent to SPC in the unconstrained case under special conditions. Even though the presented equivalence only holds under strong assumptions, it illustrates the close relationship between both methods.

In simulation studies with an exemplary LTI system with and without additive Gaussian noise, 
we showcase that the derived equivalences
also hold in practice. 
Furthermore, we investigate the non-deterministic DeePC and SPC formulation with respect to the number of recorded data sequences and the regularization term for DeePC. 
We find that SPC outperforms DeePC in the investigated cases with minor improvements of the closed-loop cost
and significant improvements in the online computation time.

In future work, we seek to further investigate the relationship between both methods 
and aim to extend the equivalence of the stochastic formulations to the constrained case. 
We also seek to compare both methods for nonlinear or time-varying systems
where the implicit re-estimation of DeePC could be beneficial.

\section*{APPENDIX}
\subsection{Proof of Lemma~\ref{lem:P_sys_response}}
\label{apdx:Proof_lemma_P_SPC}
\begin{proof}
	With the observability matrix $O_i(A,C)$ and the Toeplitz matrix $H_i(A,B,C,D)$
	\begin{equation}
		\small 
		O_i = \begin{bmatrix}
			C,\\  C  A,\\ \vdots\\  C  A^{i-1}
		\end{bmatrix},\ 
		H_i = \begin{bmatrix}
			D &  & &\\
			C  B &  D & &\\
			\vdots & & \ddots &\\
			C  A^{i-2}  B &  C A^{i-3}  B & \dots &  D
		\end{bmatrix},
	\end{equation}
	if the initial state $x_1$ is known, we can obtain $y_{T_{\text{ini}}}$ as:
	\begin{align}
		\label{eq:lemma_1_y_Tini_00}
		y_{T_{\text{ini}}}&= O_{T_{\text{ini}}} x_1 + H_{T_{\text{ini}}} u_{T_{\text{ini}}},\\
		\Leftrightarrow 
		y_{T_{\text{ini}}}
		&=
		\begin{bmatrix}
			O_{T_{\text{ini}}} & H_{T_{\text{ini}}}
		\end{bmatrix}
		\begin{bmatrix}
			x_1 \\
			u_{T_{\text{ini}}}\\
		\end{bmatrix},
		\label{eq:lemma_1_y_Tini}
	\end{align}
	Since the recorded sequences \eqref{eq:input_output_data}, \eqref{eq:X_1_matrix} are trajectories of system \eqref{eq:generic_LTI_statespace} they must satisfy:
	\begin{equation}
		Y_{T_{\text{ini}}}=
		\begin{bmatrix}
			O_{T_{\text{ini}}} & H_{T_{\text{ini}}}
		\end{bmatrix}
		\begin{bmatrix}
			X_1 \\
			U_{T_{\text{ini}}}\\
		\end{bmatrix}.
		\label{eq:lemma_1_Y_Tini}
	\end{equation}
	From \eqref{eq:lemma_1_y_Tini_00} we can also uniquely obtain $x_1$ with $y_{T_{\text{ini}}}$ and  $u_{T_{\text{ini}}}$ because Assumption~\ref{ass:lag_larger_T_ini} holds.
	This means there exists a known state-space representation of \eqref{eq:generic_LTI_statespace} with $\tilde x = \left[y_{T_\text{ini}}^{\intercal}, u_{T_\text{ini}}^{\intercal}\right]^{\intercal}$ and some system matrices $\tilde A, \tilde B, \tilde C, \tilde D$.
	Similarly as above, this allows to compute the future trajectory as:
	\begin{align}
		y_N &= \tilde O_N \tilde x + \tilde H_N u_N,\\
		\Leftrightarrow \ y_N &=
		\begin{bmatrix}
			\tilde O_N & \tilde H_N
		\end{bmatrix}	
		\begin{bmatrix}
			y_{T_\text{ini}}\\ 
			u_{T_\text{ini}}\\
			u_N
		\end{bmatrix}
		\label{eq:lemma_1_m2y_N_from_SS}
	\end{align}
	where $\tilde O_N$ is the observability matrix and $\tilde H_N$ is the Toeplitz matrix of the transformed system ($\tilde A, \tilde B, \tilde C, \tilde D$).
	The recorded sequences \eqref{eq:input_output_data} must satisfy
	\begin{equation}
		Y_N =
		\begin{bmatrix}
			\tilde O_N & \tilde H_N
		\end{bmatrix}	
		\begin{bmatrix}
			Y_{T_\text{ini}}\\ 
			U_{T_\text{ini}}\\
			U_N
		\end{bmatrix}
		= 
		\begin{bmatrix}
			\tilde O_N & \tilde H_N
		\end{bmatrix} M.
		\label{eq:lemma1_M2Y_N_fromSS}
	\end{equation}
	We multiply both sides of the equation with $M^{\dagger} M$:
	\begin{align}
		Y_N M^{\dagger} M  &=
		\begin{bmatrix}
			\tilde O_N & \tilde H_N
		\end{bmatrix} M M^{\dagger} M.
	\end{align}
	Considering the properties of the Moore-Penrose inverse ($M M^{\dagger} M = M$) we can write:
	\begin{align}
		Y_N M^{\dagger} M &= 
		\begin{bmatrix}
			\tilde O_N & \tilde H_N
		\end{bmatrix} M,
		\intertext{which yields, considering \eqref{eq:p_tilde},:}
		P^* M  &= 
		\begin{bmatrix}
			\tilde O_N & \tilde H_N
		\end{bmatrix} M
		\label{eq:lemma_1_Y_N_equal_PM}
	\end{align}
	We thus have that the projection of $M$ with $P^*$ is equivalent to the projection of $M$ with the system matrices $\tilde O_N$ and $\tilde H_N$. From \eqref{eq:lemma1_M2Y_N_fromSS}, we see that:
	\begin{equation}
		Y_N = P^*M,
	\end{equation}
	which also holds for all linear combinations of the columns of $M$ and $Y_N$ with $\alpha \in \mathbb{R}^T$:
	\begin{align}
		Y_N \alpha  &= P^*M \alpha &\forall  \alpha \in \mathbb{R}^T\\
		\Leftrightarrow y_N & = P^* 		
		\begin{bmatrix}
			u_{T_\text{ini}}\\
			u_{N}\\
			y_{T_\text{ini}}\\
		\end{bmatrix}
		&\forall \begin{bmatrix}
			y_{T_\text{ini}} \\
			u_{T_\text{ini}}\\
			u_{N}\\
		\end{bmatrix} \in \text{colspan}(M).
		\label{eq:lemma_1_y_N=Pb_b_in_colspan_M}
	\end{align}	
	We thus have that any 
	$
	\begin{bmatrix}
		y_{T_\text{ini}}^{\intercal},
		u_{T_\text{ini}}^{\intercal},
		u_{N}^{\intercal}
	\end{bmatrix}^{\intercal} \in \text{colspan}(M)
	$
	together with the resulting $y_N$ computed via \eqref{eq:multi_step_ahed_prediction} is a sequence of system \eqref{eq:generic_LTI_statespace}. 
	To show that no other sequences
	$
	\begin{bmatrix}
		y_{T_\text{ini}}^{\intercal},
		u_{T_\text{ini}}^{\intercal},
		u_{N}^{\intercal}
	\end{bmatrix}^{\intercal} \notin \text{colspan}(M)
	$
	are trajectories of system \eqref{eq:generic_LTI_statespace}, we introduce:
	\begin{equation}
		P^* = \begin{bmatrix}
			P_{y_{\text{ini}}}^*, P_{u_{\text{ini}}}^*, P_{u_N}^*
		\end{bmatrix}
		\label{eq:lemma_1_P_splitted}
	\end{equation}
	With \eqref{eq:lemma_1_Y_Tini}, \eqref{eq:lemma_1_Y_N_equal_PM} and \eqref{eq:lemma_1_P_splitted}, we can write:
	\begin{equation}
		Y_N = \begin{bmatrix}
			P_{y_{\text{ini}}}^* O_{T_{\text{ini}}}&
			P_{y_{\text{ini}}}^* H_{T_{\text{ini}}} + P_{u_{\text{ini}}}^*&
			P_{u_N}^*
		\end{bmatrix}
		\begin{bmatrix}
			X_1\\
			U_{T_\text{ini}}\\
			U_N
		\end{bmatrix}.
	\end{equation}
	By Assumption~\ref{ass:U_L_DeePC} and Assumption~\ref{ass:X_1_full_rank} the matrix
	$
	\begin{bmatrix}
		X_1^{\intercal}&
		U_{T_{\text{ini}}}^{\intercal}&
		U_N
	\end{bmatrix}^{\intercal}
	$
	has full rank and thus spans the space of all initial states $x_1 \in \mathbb{R}^n$ and sequences $u_{T_{\text{ini}}}\in \mathbb{R}^{T_{\text{ini}}m}$, $u_N\in \mathbb{R}^{Nm}$ for which all possible sequences $y_{T_{\text{ini}}}\in \mathbb{R}^{T_{\text{ini}}p}$ according to \eqref{eq:lemma_1_y_Tini} and $y_N\in \mathbb{R}^{Np}$ according to \eqref{eq:multi_step_ahed_prediction} are obtained.
	This shows that 
	$
	\begin{bmatrix}
		y_{T_\text{ini}}^{\intercal},
		u_{T_\text{ini}}^{\intercal},
		u_{N}^{\intercal},
	\end{bmatrix}^{\intercal} \in \text{colspan}(M)
	$
	spans the subspace of all possible sequences of \eqref{eq:generic_LTI_statespace} for which \eqref{eq:lemma_1_y_N=Pb_b_in_colspan_M} and thus \eqref{eq:multi_step_ahed_prediction} holds.
\end{proof}

\subsection{Proof of Result~\ref{res:v_opt_DeePC}}
\label{apdx:Proof_v_opt_DeePC}
We state the Lagrangian for problem~\eqref{eq:DeePC_noise_alt}:
\begin{equation}
	\label{eq:lagrangian_DeePC}
	\begin{gathered}
	\mathcal{L} = \frac{1}{2}g^{\intercal} (\lambda_g I) g + \frac{1}{2}y_N^{\intercal} \tilde Q y_N + \frac{1}{2}v^{\intercal} V v \\
	-\mu_1^{\intercal}
	(y_N - \tilde Y_N g)
	-\mu_2^{\intercal}
	(
	\tilde M g - b - v
	) 
	\end{gathered}
\end{equation}
The first order conditions of optimality for problem \eqref{eq:DeePC_noise_alt} can be written in terms of the Lagrangian as:
\begin{align}
	\label{eq:lagr_DeePC_01}
	\nabla_{y_N} \mathcal{L} = y_N^{\intercal}\tilde Q - \mu_1^{\intercal}  &= 0\\
	\label{eq:lagr_DeePC_02}
	\nabla_{v} \mathcal{L} =v^{\intercal}V +\mu_2^{\intercal} &= 0\\
	\label{eq:lagr_DeePC_03}
	\nabla_{g} \mathcal{L} =g^{\intercal}(\lambda_gI) +\mu_1^{\intercal}\tilde Y_N - \mu_2^{\intercal}\tilde M &= 0\\
	\label{eq:lagr_DeePC_04}
	\nabla_{\mu_1} \mathcal{L} = -y_N^{\intercal}  + g^{\intercal}\tilde Y_N^{\intercal}&=0\\
	\label{eq:lagr_DeePC_05}
	\nabla_{\mu_2} \mathcal{L} = -g^{\intercal}\tilde M^{\intercal}+b^{\intercal}+v^{\intercal}&=0
\end{align}
Inserting \eqref{eq:lagr_DeePC_01} and \eqref{eq:lagr_DeePC_02} into \eqref{eq:lagr_DeePC_03} yields:
\begin{equation}
	\label{eq:lagr_DeePC_06}
	g^{\intercal} (\lambda_gI) + y_N^{\intercal}\tilde Q \tilde Y_N + v^{\intercal}V\tilde M = 0.
\end{equation}
Next we insert \eqref{eq:lagr_DeePC_04} and \eqref{eq:lagr_DeePC_05} in \eqref{eq:lagr_DeePC_06} and rearrange, which yields:
\begin{equation}
	\label{eq:lagr_DeePC_07}
	g = (\lambda_g I + \tilde M^{\intercal}V\tilde M + \tilde Y_N^{\intercal}\tilde Q \tilde Y_N)^{-1}\tilde M^{\intercal}Vb.
\end{equation}
The inverse in \eqref{eq:lagr_DeePC_07} always exists because $\tilde M^{\intercal}V\tilde M$ and $\tilde Y_N^{\intercal}\tilde Q \tilde Y_N$ 
are positive semi-definite and $\lambda_g I$ is positive definite. 
With \eqref{eq:lagr_DeePC_05} we obtain:
\begin{equation}
	v^* = \tilde M (\lambda_g I + \tilde M^{\intercal}V\tilde M + \tilde Y_N^{\intercal}\tilde Q \tilde Y_N)^{-1}\tilde M^{\intercal}Vb-b
\end{equation}

\subsection{Proof of Result~\ref{res:v_opt_SPC}}
\label{apdx:Proof_v_opt_SPC}
\begin{proof}
We state the Lagrangian for problem~\eqref{eq:SPC_noise_alt}:
\begin{equation}
	\label{eq:lagrangian_SPC}
	\mathcal{L} = \frac{1}{2}y_N^{\intercal} Q y_N + \frac{1}{2}v^{\intercal} V v -\mu^{\intercal}
	(
	y_N-\tilde{P}b-\tilde{P}b
	) 
\end{equation}
The first order conditions of optimality for problem \eqref{eq:SPC_noise_alt} can be written in terms of the Lagrangian as:
\begin{align}
	\label{eq:lagr_SPC_01}
	\nabla_{y_N} \mathcal{L} = y_N^{\intercal}\tilde Q - \mu^{\intercal}  &= 0\\
	\label{eq:lagr_SPC_02}
	\nabla_{v} \mathcal{L} =v^{\intercal}V +\mu^{\intercal} \tilde P &= 0\\
	\label{eq:lagr_SPC_03}
	\nabla_{\mu} \mathcal{L} = -y_N^{\intercal} + v^{\intercal}\tilde P^{\intercal}+b^{\intercal} \tilde P^{\intercal}&=0
\end{align}
Inserting \eqref{eq:lagr_SPC_03} in \eqref{eq:lagr_SPC_01} and the resulting expression in \eqref{eq:lagr_SPC_02}
yields:
\begin{equation}
	\label{eq:res_v_opt_SPC_final_v_opt}
	v^* = -(V+\tilde P^{\intercal}Q \tilde P)^{-1}\tilde P^{\intercal} \tilde Q \tilde Pb
\end{equation}
after rearranging. The inverse in \eqref{eq:res_v_opt_SPC_final_v_opt} always exists, because $\tilde P^{\intercal}\tilde Q \tilde P$ is positive semi-definite 
and by assumption $V$ is positive definite.
\end{proof}

\section*{ACKNOWLEDGMENT}
The authors want to express their sincere gratitude to the reviewers for their extensive and constructive feedback.
 It has lead to many interesting discussions, new thoughts and important improvements of this work.

\bibliography{IEEEabrv,2021_ECC_Paper_DeePC_Literature}

\end{document}